\documentclass[conference]{IEEEtran}

\usepackage{graphicx}
\usepackage{latexsym}
\usepackage{amsfonts}
\usepackage{amssymb}
\usepackage{amsbsy}
\usepackage{amsmath}
\usepackage{multicol}
\usepackage{float}
\usepackage{subfigure}
\usepackage{algorithm}
 \usepackage[dvips]{epsfig}

\begin{document}

\title{Asymmetric--valued Spectrum Auction and Competition in Wireless Broadband Services}

\author{\IEEEauthorblockN{Sang Yeob Jung$^\dag$, Seung Min Yu$^\ddag$, and Seong-Lyun
Kim$^\dag$}\\
\IEEEauthorblockA{$^\dag$School of Electrical and Electronic Engineering, Yonsei University\\
50 Yonsei-Ro, Seodaemun-Gu, Seoul 120-749, Korea\\
$^\ddag$Samsung Electronics, Samsung-Ro, Yeongtong-Gu, Gyeonggi-do, 443-742, Korea\\
Email: \{syjung, smyu, slkim\}@ramo.yonsei.ac.kr}}

\maketitle

\begin{abstract}
We study bidding and pricing competition between two spiteful mobile
network operators (MNOs) with considering their existing spectrum holdings.
Given asymmetric-valued spectrum blocks are auctioned off to them via a first-price sealed-bid auction, we
investigate the interactions between two spiteful MNOs and users as a three-stage
dynamic game and characterize the dynamic game's equilibria. We show
an asymmetric pricing structure and different market share between two spiteful MNOs.
Perhaps counter-intuitively,
our results show that the MNO who acquires the less-valued spectrum block
always lowers his service price despite providing double-speed LTE service to users.
We also show that the MNO who acquires the high-valued spectrum block, despite charing a higher price,
still achieves more market share than the other MNO. We further show that the competition between two MNOs leads to some loss of their revenues.
By investigating a cross-over point at which the MNOs' profits are switched,
it serves as the benchmark of practical auction designs.
\end{abstract}

\section{Introduction}

Due to the exploding popularity of all things wireless, the demand
for wireless data traffic increases dramatically. According to a
Cisco report, global mobile data traffic will increase 13-fold
between 2012 and 2017 \cite{R_Cisco:2013}. This dramatic demand puts
on pressure on mobile network operators (MNOs) to purchase more spectrum.
However, wireless spectrum is a scarce resource for mobile services.
Even if the continued innovations in technological progress relax
this constraint as it provides more capacity and higher
quality of service (QoS), the shortage of spectrum is still the
bottleneck when the mobile telecommunications industry is moving
toward wireless broadband services \cite{Kleeman:2011}.

To achieve a dominant position for future wireless services, thus,
it is significant how new spectrum is allocated to MNOs. Since the
spectrum is statically and infrequently allocated to an MNO, there
has been an ongoing fight over access to the spectrum. In South
Korea, for example, the Korea Communications Commission (KCC) planed
to auction off additional spectrum in both 1.8 GHz and 2.6 GHz
bands. The main issue was whether Korea Telecom (KT) acquires the
contiguous spectrum block or not. Due to the KT's existing holding
downlink 10 MHz in the 1.8 GHz band, it could immediately double the
existing Long Term Evolution (LTE) network capacity in the 1.8 GHz
band at little or no cost. This is due to the support of the
downlink up to 20 MHz contiguous bandwidth by LTE Release 8/9. To
the user side, there is no need for upgrading their handsets. LTE
Release 10 (LTE-A) can support up to 100 MHz bandwidth but this
requires the carrier aggregation (CA) technique, for which both
infrastructure and handsets should be upgraded. If KT leases the spectrum block in the 1.8
GHz band, KT might achieve a dominant position in the market. On the
other hand, other MNOs expect to make heavy investments as well as
some deployment time to double their existing LTE network capacities
compared to KT \cite{Melissa:2011}. Thus, the other MNOs requested
the government to exclude KT from bidding on the contiguous
spectrum block to ensure market competitiveness. Although we
consider the example of South Korea, this interesting but
challenging issue on spectrum allocation is not limited to South
Korea but to most countries when asymmetric-valued spectrum blocks
are auctioned off to MNOs.

Spectrum auctions are widely used by governments to allocate
spectrum for wireless communications.
Most of the existing auction literatures assume that each bidder
(i.e., an MNO) only cares about his own profit: what spectrum block
he gets and how much he has to pay \cite{P. Milgrom:1982}. Given
spectrum constraints, however, there is some evidence that a bidder
considers not only to maximize his own profit in the event that he wins the auction
but to minimize the weighted difference of his competitor's profit and his own profit in the event that he loses the auction \cite{J. Morgan_05}. This strategic concern can be
interpreted as a \textit{spite motive}, which is the preference to
make competitors worse off. Since it might increase the MNO's
relative position in the market, such concern has been observed in
spectrum auctions \cite{Brandt:2007}.

In this paper, we study bidding and pricing competition between two competing/spiteful MNOs
with considering their existing spectrum holdings.
Given that asymmetric-valued spectrum blocks are auctioned off to them,
we developed an analytical framework to investigate the interactions between two MNOs and users as a three-stage dynamic game.
In \textrm{S}tage I, two spiteful MNOs compete in a first-price sealed-bid auction.
Departing from the standard auction framework, we address the bidding behavior of the spiteful MNO.
In \textrm{S}tage II, two competing MNOs optimally set their
service prices to maximize their revenues with the newly allocated
spectrum. In \textrm{S}tage III, users decide whether to stay in
their current MNO or to switch to the other MNO for utility
maximization.

Our results are summarized as follows:
\begin{itemize}
\item \textit{Asymmetric pricing structure}: We show that two MNOs announce different equilibrium prices to the users, even providing the same quality in services to the users.
\item \textit{Different market share}: We show that the market share leader, despite charging a higher price, still achieve more market share.
\item \textit{Impact of competition}: We show that the competition between two MNOs leads to some loss of their revenues.
\item \textit{Cross-over point between two MNO's profits}: We show that two MNOs' profits are switched.
\end{itemize}

The rest of the paper is organized as follows: Related works are discussed in \textrm{S}ection II. The system model and three-stage dynamic game are described in \textrm{S}ection III.
Using backward induction, we analyze user responses and pricing competition in \textrm{S}ections VI and V, and bidding competition in \textrm{S}ection VI.
We conclude in Section \textrm{V}II together with some future research directions.

\section{Related Work}
In wireless communications, the competition among MNOs have been
addressed by many researchers \cite{Min Yu:2013}--\cite{Feng:2012}.
Yu and Kim \cite{Min Yu:2013} studied price dynamics among MNOs.
They also suggested a simple regulation that guarantees a
Pareto optimal equilibrium point to avoid instability and
inefficiency.
Niyato and Hossain \cite{D. Niyato_08} proposed a
pricing model among MNOs providing different services to users.
However, these works did not consider the spectrum allocation issue.
More closely related to our paper are some recent works
\cite{J.Jia_08}--\cite{Feng:2012}.
The paper \cite{J.Jia_08} studied bandwidth and price competition (i.e., Bertrand competition) among MNOs. By
taking into account MNOs' heterogeneity in leasing costs and users' heterogeneity in transmission power and channel conditions,
Duan \textit{et al}. presented a comprehensive analytical study of MNOs' spectrum leasing and pricing strategies in \cite{L. Duan_11}.
In \cite{Chen:2013}, a new allocation scheme is suggested by jointly considering MNOs' revenues and social welfare.
X. Feng \textit{et al.} \cite{Feng:2012} suggested a truthful double auction scheme for heterogeneous spectrum allocation.
None of the prior results considered MNOs' existing spectrum holdings even if the value of spectrum could be varied depending on
MNOs' existing spectrum holdings.

\section{System Model and Game Formulation}

We consider two MNOs ($i, j \in \{1,2\}$ and $i \ne j$) compete in a first-price sealed-bid
auction\footnote{It is a form of auction where two MNOs submit one
bid in a concealed fashion. The MNO with the highest bid wins and
pays his bid for the spectrum block.}, where two spectrum blocks $A$
and $B$ are auctioned off to them as shown in \textrm{F}ig. 1. Note
that $A$ and $B$ are the same amount of spectrum (i.e., 10 MHz
spectrum block). Without loss of generality, we consider only the
downlink throughput the paper. Note that both MNOs operate Frequency
Division Duplex LTE (FDD LTE) in the same area.

Due to the MNOs' existing spectrum holdings (i.e., each MNO secures
10 MHz downlink spectrum in the 1.8 GHz band), the MNOs put values
on spectrum blocks $A$ and $B$ asymmetrically. If MNO $i$ leases $A$,
twice (2x) improvements in capacity over his existing LTE network
capacity are directly supported to users. In Third Generation
Partnership Project (3GPP) LTE Release 8/9, LTE carriers can support
a maximum bandwidth of 20 MHz for both in uplink and downlink,
thereby allowing for MNO $i$ to provide double-speed LTE service to
users without making many changes to the physical layer structure of
LTE systems \cite{Sesia:2009}. On the other hand, MNO $j$ who leases
$B$ should make a huge investment to double the capacity after some
deployment time $T_1$. Without loss of generality, we assume that MNO $i$ leases $A$.

To illustrate user responses, we define the following
terms as follows.

\vskip 10pt \noindent {\bf Definition 1.} (Asymmetric phase) {\it
Assume that MNO $j$ launches double-speed LTE service at time $T_1$.
When $0\leq t \leq T_1$, we call this period asymmetric phase due to
the different services provided by MNOs $i$ and $j$. }

\noindent {\bf Definition 2.} (Symmetric phase) {\it Assume that
$T_2$ denotes the expiration time for the MNOs' new spectrum rights.
When $T_1< t \leq T_2$, we call this period symmetric phase because
of the same services offered by MNOs $i$ and $j$.}
\begin{figure}[t]
\centering
\includegraphics[width=3.5in]{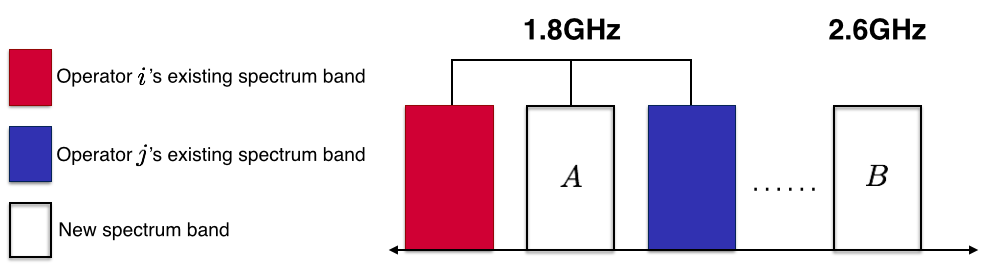}
\caption{System model for spectrum auction. Without loss of
generality, we consider only the downlink throughput the paper.}
\end{figure}

\begin{figure}[t]
\centering
\includegraphics[angle=0,width=3.4in]{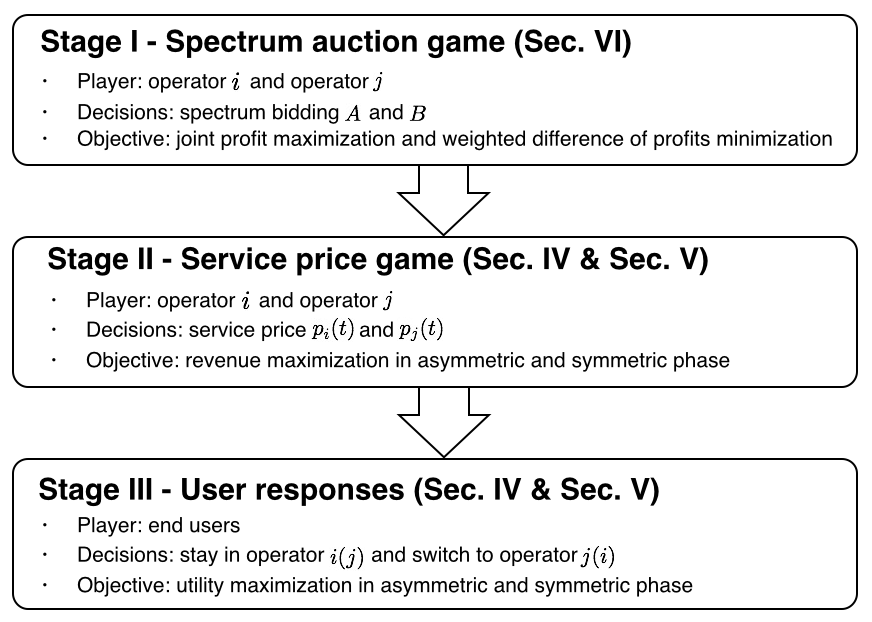}
\caption{Three stages of the dynamic game.}
\end{figure}
We investigate the interactions between two MNOs and users as a
three-stage dynamic game as shown in \textrm{F}ig. 2. In
\textrm{S}tage I, two spiteful MNOs compete in a first-price
sealed-bid auction where asymmetric-valued spectrum blocks $A$ and
$B$ are auctioned off to them. The objective of each MNO is maximizing
his own profit when $A$ is assigned to him, as well as minimizing the weighted difference of his competitor's profit
and his own profit when $B$ is allocated to him.
In \textrm{S}tage II, two competing MNOs optimally announce their service prices to maximize
their revenues given the result of \textrm{S}tage I. The analysis is
divided into two phases: asymmetric phase and symmetric phase. In
\textrm{S}tage III, users determine whether to stay in their current
MNO or to switch to the new MNO for utility maximization.
To predict the effect of spectrum allocation, we solve this three-stage dynamic game by applying the concept of backward
induction, from \textrm{S}tage III to \textrm{S}tage I.

\section{User Responses and Pricing Competition \\in Asymmetric Phase}

\subsection{User Responses in \textrm{S}tage III in Asymmetric Phase}
Each user subscribes to one of the MNOs based on his or her MNO
preference. Let us assume that MNOs $i$ and $j$ provide same
quality in services to the users so they have the same reserve
utility $u_o$ before spectrum auction. Each MNO initially has
50\% market share and the total user population is normalized to 1.

In asymmetric phase, the users in MNOs $i$ and $j$ obtain different
utilities, i.e.,
\begin{eqnarray}
u_i(t)=(1+\eta)u_o, \;u_j(t)=u_o, \;0\leq t \leq T_1.
\end{eqnarray} \noindent
where $\eta \in (0,1)$ is a user sensitivity parameter to the
double-speed LTE service than existing one. It means that users care more about the data rate as $\eta$ increases.
The users in MNO $j$ have more incentive to switch to MNO $i$ as $\eta$ increases. When
they decide to change MNO $i$, however, they face switching costs, the disutility that a user experiences from
switching MNOs. In the case of higher
switching costs, the users in MNO $j$ have less incentive to switch.
The switching cost varies among users and discounts over time. To
model such users' time-dependent heterogeneity, we assume that the
switching cost is heterogeneous across users and uniformly
distributed in the interval $[0,e^{ - \lambda t} ]$ at $t\geq0$,
where $\lambda$ denotes the discount rate \cite{Mengze:2006}.
This is due to the fact that the pays for the penalty of terminating contract with operators
decrease as time passes.

Now let us focus on how users churn in asymmetric phase. A user $k$
in MNO $j$, with switching cost, $s_k(t)$, observes the prices
charged by MNOs $i$ and $j$ ($p_i(t)$ and $p_j(t)$). A user $k$ in MNO $j$
will switch to MNO $i$ if and only if
\begin{equation}
u_j(t) - p_j(t) \leq u_i(t) - p_i(t) - s_k(t), \;0\leq t \leq T_1.
\end{equation}
Thus the mass of switching users from MNO $j$ to $i$ is
\setlength\arraycolsep{-1pt} \medmuskip=-2mu \thickmuskip=-0.5mu
\begin{eqnarray}
Q_{j \to i} (t) = \frac{1} {2}\int_0^{\eta u_o  + p_j (t) - p_i (t)}
{e^{\lambda t} ds}  = \frac{{e^{\lambda t} \left( {\eta u_o  + p_j
(t) - p_i (t)} \right)}}
{2}, \nonumber \\
\end{eqnarray}
where $s$ is a uniform $(0,1)$ random variable and $\frac{1}{2}$ denotes the initial market share.

Since the market size is normalized to one, each MNO's market share
in asymmetric phase is as follows:
\setlength\arraycolsep{2pt} \thickmuskip=1mu \medmuskip=1mu
\begin{eqnarray}
Q_i (t) = \frac{1}{2} + Q_{j \to i}(t),\; Q_j(t)= \frac{1}{2} - Q_{j \to i}(t).
\end{eqnarray}

\subsection{Pricing Competition in \textrm{S}tage II in Asymmetric Phase}
Given users' responses (4), MNOs $i$ and $j$ set their
service prices $p_i^ *  (t)$ and $p_j^ *  (t)$ to maximize their revenues,
respectively, i.e.,
\begin{eqnarray}
p_i^ *  (t) = \mathop {\arg \max }\limits_{p_i (t)} {\text{ }}p_i
(t)Q_i (t),{\text{ }} i,j \in \{ 1,2\} \;\;\;\;\;{\rm{and}}\;\;\; \;\;i \ne j.
\end{eqnarray}
The Nash equilibrium in this pricing game is described in the following
proposition.

\vskip 10pt \noindent {\bf Proposition 1.}  {\it When $0\leq t\leq
T_1$ and $\eta u_o<e^{-\lambda t}$, there exists a unique Nash
equilibrium, i.e.,
\begin{eqnarray}
p_i^ *  (t) = e^{ - \lambda t}  + \frac{1} {3}\eta u_o ,{\text{
}}p_j^ *  (t) = e^{ - \lambda t}  - \frac{1} {3}\eta u_o.
\end{eqnarray}}
\vskip 10pt \noindent {\bf Proof.} In asymmetric phase, two
competing MNOs try to maximize their revenues $r_i(t)$ and
$r_j(t)$, respectively, given users' responses, i.e.,
\begin{eqnarray}
\mathop {\max }\limits_{p_i (t)} r_i (t) = p_i (t)Q_i (t),{\text{
}}i,j \in \{ 1,2\} \;\;\;\;\;{\rm{and}}\;\;\; \;\;i \ne j. \nonumber
\end{eqnarray}
A Nash equilibrium exists by satisfying and solving the following
first order conditions with respect to $p_i(t)$ and $p_j(t)$, i.e.,
\begin{eqnarray}
\frac{{\partial r_i (t)}} {{\partial p_i (t)}} = \frac{{1 + \left(
{\eta u_o  + p_j (t) - 2p_i (t)} \right)e^{  \lambda t} }} {2} = 0,
\nonumber
\end{eqnarray}
\begin{eqnarray}
\frac{{\partial r_j (t)}} {{\partial p_j (t)}} = \frac{{1 - \left(
{\eta u_o  - p_i (t) + 2p_j (t)} \right)e^{  \lambda t} }} {2} = 0.
\nonumber
\end{eqnarray}
\hfill $\blacksquare$

Proposition 1 shows two MNOs' equilibrium prices in asymmetric
phase. Intuitively, $p_i^ *  (t)$ increases as $\eta$ increases.
With larger $\eta$, users care more about the data rate. Thus, MNO $i$
increases his service price to obtain more revenue. On the other
hand, $p_j^* (t)$ decreases as $\eta$ increases. It means that MNO
$j$ tries to sustain the revenue margin by lowering the service price and
holding onto market share. An interesting observation is that both
MNOs decrease their service prices as $t$ increases. Due to the
discount factor ($\lambda$), the users in MNO $j$ are not locked-in
and tries to maximize their utilities by churning to MNO $i$ as
switching costs decrease over time. Therefore, MNO $i$ lowers his
service price to maximize his revenue, which forces MNO $j$ to
decrease the service price. This phenomenon is consistent with the
previous results \cite{Min Yu:2013}, \cite{Mengze:2006} in that the reduction of
switching costs intensifies the price-down competition between two MNOs. If $\eta
u_o  > e^{ - \lambda t}$, then all users in MNO $j$ churn to MNO $i$. However, it is an unrealistic feature of the mobile
telecommunication industry so we add the constraint $\eta u_o  < e^{
- \lambda t}$.

Next we will show how each MNO's market share changes in asymmetric
phase. Inserting the equilibrium prices (6) into (4), each MNO's
market share can be calculated as follows:
\begin{eqnarray}
Q_i(t)  = \frac{{3 + \eta u_o e^{\lambda t} }} {6},{\text{
}}Q_j(t) = \frac{{3 - \eta u_o e^{\lambda t} }} {6},{\text{
}} 0\leq t \leq T_1.
\end{eqnarray}
Intuitively, MNO $i$ takes MNO $j$'s market share more as $t$
increases or $\eta$ increases. To hold onto or take MNO $i$'s market
share, the time to launch double-speed LTE service $T_1$ is of great
importance to MNO $j$.

When MNO $j$ launches double-speed LTE service at time $T_1$, each
MNO's total revenue in asymmetric phase is given by
\setlength\arraycolsep{-2pt} \medmuskip=-0.5mu
\begin{eqnarray}
r_i(T_1) = \int_0^{T_1 } {r_i (t)dt}  = \frac{{1 - e^{ -
\lambda T_1 } }} {{2\lambda }} - \frac{{1 - e^{\lambda T_1 } }}
{{18\lambda }}\left( {\eta u_0 } \right)^2  + \frac{1} {3}\eta u_0
T_1, \nonumber
\end{eqnarray}
\setlength\arraycolsep{-1pt}
\begin{eqnarray}
r_j (T_1) = \int_0^{T_1 } {r_j (t)dt}  = \frac{{1 - e^{ -
\lambda T_1 } }} {{2\lambda }} - \frac{{1 - e^{\lambda T_1 } }}
{{18\lambda }}\left( {\eta u_0 } \right)^2  - \frac{1} {3}\eta u_0
T_1.
\nonumber \\
\end{eqnarray}
Similar to the analysis of market share, Equation (8) shows that MNO $j$ should launch double-speed LTE service
as quickly as possible to narrow the revenue gap between MNO $i$ and MNO $j$ (see the last term of the revenues (8)).

\section{User Responses and Pricing Competition \\in Symmetric Phase}
\subsection{User Responses in Stage III in Symmetric Phase}
Since MNO $j$ launches double-speed LTE service in symmetric phase,
we assume that the users in MNOs $i$ and $j$ obtain
same utility, i.e.,
\begin{eqnarray}
u_i(t)=u_j(t)=(1+\eta)u_o, \;T_1<t\leq T_2.
\end{eqnarray}

For better understanding of user responses in symmetric phase, we first discuss
the effect of switching costs on market competition. Given the same services offered by two MNOs,
an MNO's current market share plays an important role in determining its price strategy.
Each MNO faces a trade-off between a low price to
increase market share, and a high price to harvest profits
by exploiting users' switching costs. The following \textrm{L}emma
examines this trade-off and characterizes each MNO's price strategy,
which is directly related to user responses in symmetric phase.
\vskip 10pt \noindent {\bf Lemma 1.} {\it In a competitive market
with switching costs, the market share leader (i.e., MNO $i$) charges a
high price to exploit its current locked-in users while the marker
share followers (i.e., MNO $j$) charge low prices to increase market
share for revenue maximization, respectively, given the same
services offered by them.} \vskip 10pt \noindent {\bf Proof.} We
prove \textrm{L}emma 1 by contradiction. Suppose that MNO $j$ charges
a higher price than MNO $i$ (i.e., $p_i(t) < p_j(t), T_1 < t \leq
T_2)$.
The mass of switching users from MNO $j$ to $i$ is
\setlength\arraycolsep{-1pt} \medmuskip=-1mu \thickmuskip=0mu \thinmuskip=0mu
\begin{eqnarray}
Q_{j \to i} (t) =Q_j (T_1) \int_0^{p_j(t) - p_i(t)}
{e^{\lambda
t} ds}  = \left( {p_j (t) - p_i (t)} \right)Q_j(T_1 )e^{\lambda t}, \nonumber \\
\end{eqnarray}
where $Q_j(T_1)=\frac{{3 - \eta u_o e^{\lambda T_1} }} {6} $
is the market share of MNO $j$ at the end of asymmetric phase. Then,
each MNO's market share is given by
\begin{eqnarray}
Q_i (t) = Q_i (T_1 ) - \left( {p_j (t) - p_i (t)}
\right)Q_j(T_1 ) e^{\lambda t}  , \nonumber
\end{eqnarray}
\begin{eqnarray}
Q_j (t) = Q_j (T_1 ) \left( {1 - \left( {p_j (t) - p_i (t)}
\right) e^{\lambda t} } \right),
\end{eqnarray}
where $Q_i (T_1)=\frac{{3 + \eta u_o e^{\lambda T_1} }} {6} $
is the market share of MNO $i$ at the end of asymmetric phase.
Following the same steps of the \textrm{P}roposition 1, we can find
the Nash equilibrium by satisfying and solving the following first
order conditions with respect to $p_i(t)$ and $p_j(t)$, i.e.,
\begin{eqnarray}
\frac{{\partial r_i (t)}} {{\partial p_i (t)}} =Q_i (T_1 ) +
\left( {p_j (t) - 2p_i (t)} \right)Q_j (T_1 )e^{\lambda t}  = 0
, \nonumber
\end{eqnarray}
\begin{eqnarray}
\frac{{\partial r_j (t)}} {{\partial p_j (t)}} =Q_j (T_1
)\left[ {1 - \left( {2p_j (t) - p_i (t)} \right)e^{\lambda t} }
\right] = 0,\nonumber
\end{eqnarray}
which yields the solution given as follows
\begin{eqnarray}
p_1^ *  (t) = \left( {\frac{{9 + \eta u_0 e^{\lambda T_1 } }} {{9 -
3\eta u_0 e^{\lambda T_1 } }}} \right)e^{ - \lambda t} ,\;\;p_j^ *
(t) = \left( {\frac{{9 - \eta u_0 e^{\lambda T_1 } }} {{9 - 3\eta
u_0
e^{\lambda T_1 } }}} \right)e^{ - \lambda t}.\nonumber \\
\end{eqnarray}
Thus, this contradicts to our assumption, completing the
proof. \hfill$\blacksquare$

With \textrm{L}emma 1, let us illustrate the process of user
churn in symmetric phase.
The mass of switching users from MNO $i$ to $j$ is
\setlength\arraycolsep{-2pt} \medmuskip=-1mu
\begin{eqnarray}
Q_{i \to j} (t) =Q_1 (T_1 ) \int_0^{p_i (t) - p_j (t)}
{e^{\lambda
t} ds}  = \left( {p_i (t) - p_j (t)} \right)Q_i (T_1 )e^{\lambda t}, \nonumber \\
\end{eqnarray}
where $Q_i  (T_1)=\frac{{3 + \eta u_o e^{\lambda T_1} }} {6}$
is the market share of MNO $i$ at the end of asymmetric phase. Then
each MNO's market share in symmetric phase is
given by
\begin{eqnarray}
Q_i (t) = Q_i  (T_1 ) \left( {1 - \left( {p_i (t) - p_j (t)}
\right) e^{\lambda t} } \right), \nonumber
\end{eqnarray}
\begin{eqnarray}
Q_j (t) = Q_j (T_1 ) + Q_i (T_1 )\left( {p_i (t) - p_j
(t)} \right)e^{\lambda t},
\end{eqnarray}
where $Q_j (T_1)=\frac{{3 - \eta u_o e^{\lambda T_1} }} {6} $
is the market share of MNO $j$ at the end of asymmetric phase.

\subsection{Pricing Competition in Symmetric Phase in \textrm{S}tage II}

As noted in \textrm{L}emma 1, MNO $j$ charges a lower price than MNO $i$ in symmetric phase.
Following the same procedure (5), the Nash equilibrium is described in the following proposition. \vskip
10pt \noindent {\bf Proposition 2.} {\it When $T_1< t\leq T_2$,
there exists a unique Nash equilibrium, i.e.,
\setlength\arraycolsep{1pt} \medmuskip=-0.5mu \thickmuskip=0mu
\begin{eqnarray}
p_i^ *  (t) = \left( {\frac{{9 + \eta u_0 e^{\lambda T_1 } }} {{9 +
3\eta u_0 e^{\lambda T_1 } }}} \right)e^{ - \lambda t} ,\;\;p_j^ *
(t) = \left( {\frac{{9 - \eta u_0 e^{\lambda T_1 } }} {{9 + 3\eta
u_0
e^{\lambda T_1 } }}} \right)e^{ - \lambda t}.\nonumber \\
\end{eqnarray}}
\vskip 10pt \noindent {\bf Proof.} Following the same steps of the
\textrm{P}roposition 1, a Nash equilibrium exists by satisfying and
solving the following first order conditions with respect to
$p_i(t)$ and $p_j(t)$, i.e.,
\begin{eqnarray}
\frac{{\partial r_i (t)}} {{\partial p_i (t)}} = Q_i (T_1
)\left[ {1 - \left( {2p_i (t) - p_j (t)} \right)e^{\lambda t} }
\right] = 0 , \nonumber
\end{eqnarray}
\begin{eqnarray}
\frac{{\partial r_j (t)}} {{\partial p_j (t)}} = Q_j (T_1 ) -
\left( {2p_j (t) - p_i (t)} \right)Q_i (T_1 )e^{\lambda t}  =
0. \nonumber
\end{eqnarray}
\hfill $\blacksquare$

\begin{figure}[t]
\centering
\includegraphics[angle=0,width=3.4in]{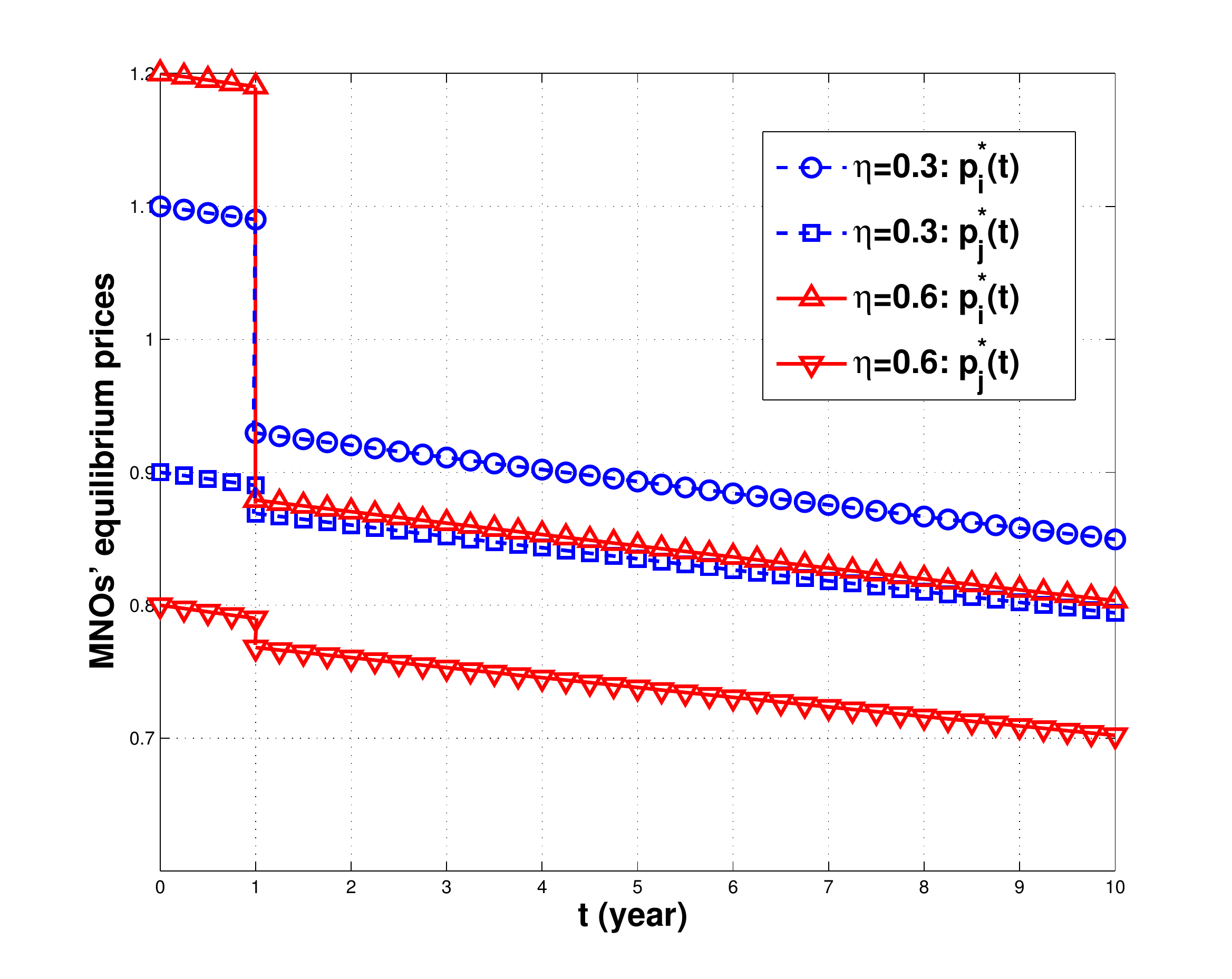}

\caption{MNOs' equilibrium prices in asymmetric and symmetric phase
under two different user sensitivities ($\eta=0.3$, $\eta=0.6$).
Other parameters are $u_o=1$, $\lambda=0.01$, $T_1=1$, and
$T_2=10$.}
\end{figure}

\begin{figure}[t]
\centering
\includegraphics[angle=0,width=3.4in]{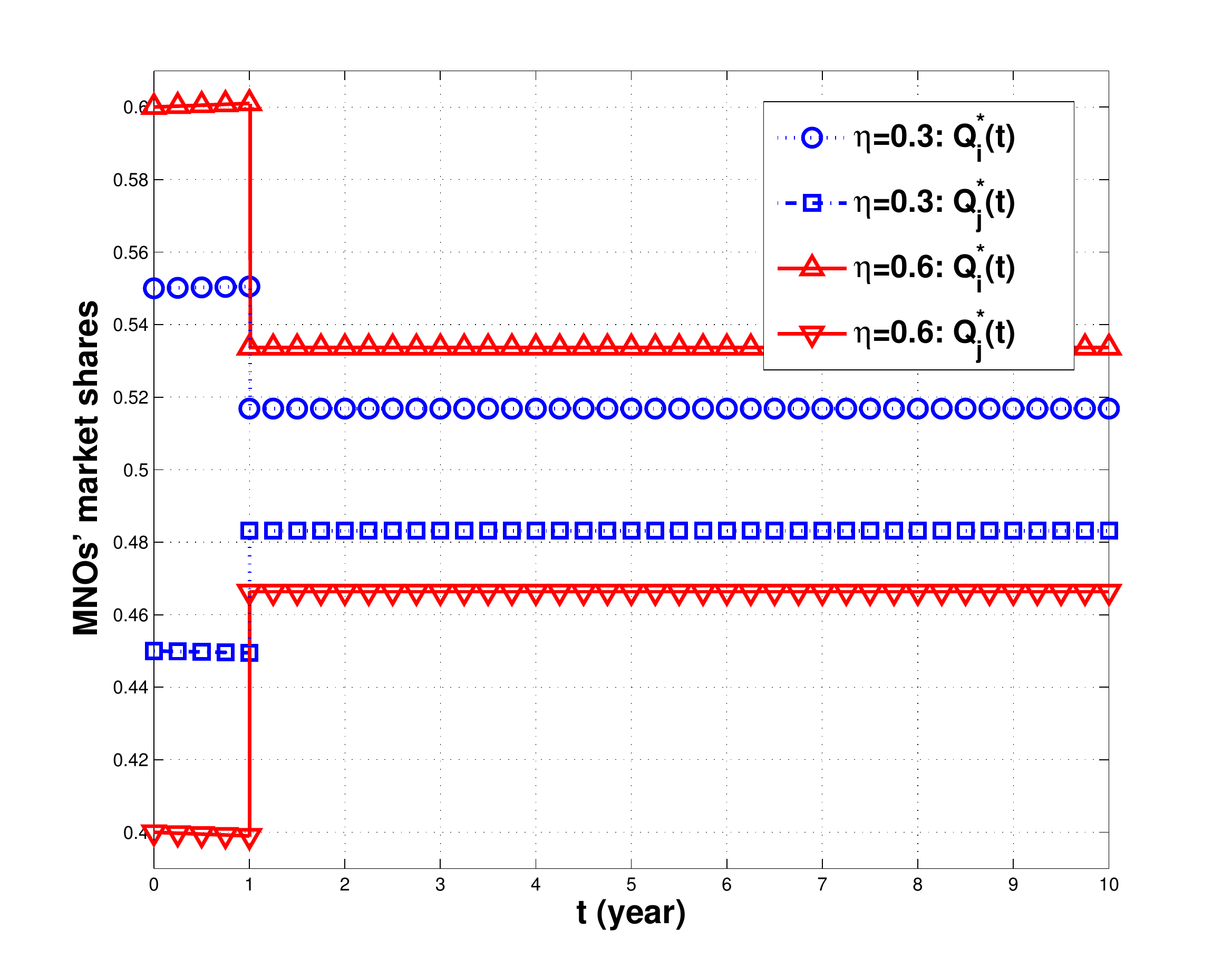}
\caption{User responses in asymmetric and symmetric phase under two
different user sensitivities ($\eta=0.3$, $\eta=0.6$). Other
parameters are $u_o=1$, $\lambda=0.01$, $T_1=1$ and $T_2=10$.}
\end{figure}

\textrm{P}roposition 2 states the MNOs' equilibrium prices in
symmetric phase. As described in \textrm{L}emma 1, MNO $i$, the
market share leader announces a higher service price up to $\frac{{2\eta u_o e^{\lambda T_1 } }} {{9 +
3\eta u_o e^{\lambda T_1 } }}e^{ - \lambda t}$ than MNO $j$.

To further investigate the effect of
competition under the same quality in services, let us calculate each MNO's falling
price level in the neighborhood of the point $T_1$. From (6) and (15), each MNO's
falling price level (i.e., $\varphi _i (T_1 )$ and $\varphi _j (T_1
)$) is \setlength\arraycolsep{4pt} \medmuskip=0mu
\thickmuskip=0mu \thinmuskip=0mu
\begin{eqnarray}
\varphi _i (T_1 )=\mathop {\lim }\limits_{\varepsilon  \to 0} \left(
{p_i^ *  (T_1-\varepsilon ) - p_i^ *  (T_1  + \varepsilon )} \right)
= \frac{{\eta u_0 \left( {5 + \eta u_0 e^{\lambda T_1 } } \right)}}
{{9 + 3\eta u_0 e^{\lambda T_1 } }}, \nonumber
\end{eqnarray}
\begin{eqnarray}
\varphi _2 (T_1 )=\mathop {\lim }\limits_{\varepsilon  \to 0} \left(
{p_j^ *  (T_1-\varepsilon ) - p_j^ *  (T_1  + \varepsilon )} \right)
= \frac{{\eta u_0 \left( {1 - \eta u_0 e^{\lambda T_1 } } \right)}}
{{9 + 3\eta u_0 e^{\lambda T_1 } }}.
\end{eqnarray}
Because $0<\eta u_o  < e^{ - \lambda T_1 }$ , MNOs $i$ and $j$ always
decrease their prices up to $\varphi _i (T_1 )$ and $\varphi _j (T_1)$ at the starting point of the symmetric phase, respectively.
Perhaps counter-intuitively, it shows that MNO $j$ always lowers his
price despite launching double-speed LTE service at the starting
point of the symmetric phase. It can be interpreted as follows.
Since MNO $j$ loses his market share in asymmetric phase, MNO $j$
attempts to maximize his revenue by lowering his service price and
increasing his market share, which forces MNO $i$ to drop the service price at the same
time. This means that the MNOs' competition under the same quality
in services lead to some loss of their revenues, which, known as a
\textit{price war}, is
consistent with our previous work \cite{Min Yu:2013}. \textrm{F}ig. 3
shows $p_i^*(t)$ and $p_j^*(t)$ as a function of $t$ under two
different user sensitivities ($\eta=0.3$, $\eta=0.6$). Note that MNO $i$'s falling price level is more sensitive to $\eta$.

Next we show that how each MNO's market share varies in symmetric
phase. From (14) and (15), each MNO's market share is
\begin{eqnarray}
Q_i  (t) = \frac{1} {2} + \frac{{\eta u_0 e^{\lambda T_1 } }}
{{18}}, Q_j  (t) = \frac{1} {2} - \frac{{\eta u_0 e^{\lambda T_1 } }}
{{18}}, {\text{
}} T_1 < t \leq T_2.
\end{eqnarray}
Unlike the asymmetric phase, each MNO's market share only depends on the deployment time of carrier aggregation $T_1$ in symmetric phase.
An interesting observation is that the market
share leader (i.e., MNO $i$), despite charging a higher price, still
achieves more market share up to $\frac{1} {9}\eta u_0 e^{\lambda
T_1 }$ than MNO $j$. In terms of market share,
MNO $i$ always gains a competitive advantage over MNO $j$ if MNO $j$
was forced to lease less-valued spectrum block.
This explains how critical new spectrum is allocated to the MNOs,
and how struggling they are over access to the spectrum for improving market competitiveness for future wireless services.
Fig. 4 shows user
responses as a function of $t$ under two different user
sensitivities ($\eta=0.3$, $\eta=0.6$).

If the new spectrum rights expire at $t=T_2$, each MNO's total revenue
in symmetric phase is
\setlength\arraycolsep{3pt}
\medmuskip=2mu
\begin{eqnarray}
r_i (T_1, T_2 ) = \int_{T_1 }^{T_2 } {r_i  (t)dt}  = \frac{{\left(
{9 + \eta u_0 e^{\lambda T_1 } } \right)^2 }} {{54\left( {3 + \eta
u_0 e^{\lambda T_1 } } \right)}}\left( {\frac{{e^{ - \lambda T_1 } -
e^{ - \lambda T_2 } }} {\lambda }} \right), \nonumber
\end{eqnarray}
\setlength\arraycolsep{1pt}
\medmuskip=-1mu\thinmuskip=-1mu\thickmuskip=-1mu
\begin{eqnarray}
r_j  (T_1, T_2 ) = \int_{T_1 }^{T_2 } {r_j  (t)dt}  = \frac{{\left(
{9 - \eta u_0 e^{\lambda T_1 } } \right)^2 }} {{54\left( {3 + \eta
u_0 e^{\lambda T_1 } } \right)}}\left( {\frac{{e^{ - \lambda T_1 } -
e^{ - \lambda T_2 } }} {\lambda }} \right). \nonumber \\
\end{eqnarray}

\begin{figure*}[t!]
\begin{equation}
r^A(T_1,T_2)   = r_i  (T_1 ) + r_i  (T_1, T_2 ) = \frac{{1 - e^{ -
\lambda T_1 } }} {{2\lambda }} - \left( {\frac{{1 - e^{\lambda T_1 }
}} {{18\lambda }}} \right)\left( {\eta u_0 } \right)^2  + \frac{1}
{3}\eta u_0 T_1  + \frac{{\left( {9 + \eta u_0 e^{\lambda T_1 } }
\right)^2 }} {{54(3 + \eta u_0 e^{\lambda T_1 } )}}\left(
{\frac{{e^{ - \lambda T_1 }  - e^{ - \lambda T_2 } }} {\lambda }}
\right), \nonumber
\end{equation}
\begin{equation}
r^B(T_1,T_2)   = r_j  (T_1 ) + r_j  (T_1, T_2 ) = \frac{{1 - e^{ -
\lambda T_1 } }} {{2\lambda }} - \left( {\frac{{1 - e^{\lambda T_1 }
}} {{18\lambda }}} \right)\left( {\eta u_0 } \right)^2  - \frac{1}
{3}\eta u_0 T_1  + \frac{{\left( {9 - \eta u_0 e^{\lambda T_1 } }
\right)^2 }} {{54(3 + \eta u_0 e^{\lambda T_1 } )}}\left(
{\frac{{e^{ - \lambda T_1 }  - e^{ - \lambda T_2 } }} {\lambda }}
\right).
\end{equation}
\hrulefill
\end{figure*}

Using (8) and (18), we examine the two MNOs' aggregate revenues when
MNO $i$ leases $A$ and MNO $j$ leases
$B$. Each MNO's aggregate revenue at $t=T_2$ is given in (19).

When MNO $j$ decides to launch double-speed LTE service, the optimal deployment time of the carrier aggregation $T_1^\ast$ should be
studied.
The following Lemma describes the MNO $j$'s optimal deployment time.
\vskip 10pt \noindent {\bf Lemma 2.} {\it The market share followers (i.e., MNO $j$) should launch
double-speed LTE service as quickly as possible not only for maximizing their own revenues
but also for minimizing the market leader's revenue.} \vskip 10pt \noindent {\bf Proof.}
By taking the derivative of the two MNO's aggregate revenues $r^A(T_1,T_2)$ and $r^B(T_1,T_2)$ with respect to $T_1$, respectively,
it can be checked that ${{\partial r^A (T_1 ,T_2 )} \over {\partial T_1 }}>0$ and ${{\partial r^B (T_1 ,T_2 )} \over {\partial T_1 }}<0$. We omit the details of the derivations here.
\hfill$\blacksquare$

\textrm{L}emma 2 states that the revenue of MNO $j$ is strictly decreasing over $T_1$
while the reverse is for MNO $i$. To gain more insight into the effect of the allocation
of asymmetric-valued spectrum blocks, let us define the revenue gain as follows:
\begin{eqnarray}
r_{gain}  = \frac{{r^A(T_1,T_2)  }} {{r^B(T_1,T_2)  }}.
\end{eqnarray}
\textrm{F}ig. 5 shows the revenue gain as a function of $\eta$ under
two different deployment times ($T_1=1$, $T_1=2$). As expected, the revenue
gain is strictly increasing over $T_1$ and $\eta$. In terms of $\eta$,
it can be checked directly by following the same steps of the \textrm{L}emma 2.
Such result explains why each MNO should spitefully bid in a first-price
sealed-bid auction to achieve a dominant position or compensate the revenue gap, which we will
discuss these points in the next section.

\section{Bidding Competition in \textrm{S}tage I}

In \textrm{S}tage I, two spiteful MNOs $i$ and $j$ compete in a first-price sealed-bid auction
where asymmetric-valued spectrum blocks $A$ and $B$ are auctioned off to them.
For fair competition, each MNO is constrained to lease only one spectrum block (i.e., $A$ or $B$).
We assume that the governments set the reserve prices $c_A$ and $c_B$ to $A$ and $B$, respectively.
Note that the reserve price is the minimum price to get the spectrum block.
Since $A$ is the high-valued spectrum block, we further assume that two spiteful MNOs
are only competing on $A$ to enjoy a dominant position in the market.
MNOs $i$ and $j$ bid $A$ independently as $b_i$ and $b_j$, respectively.
In this case, $B$ is assigned to the MNO who loses in the auction as the reserve price $c_B$.
Because the MNO who leases $B$ should make huge investments to double the existing LTE network capacity
compared to the other MNO, we also assume the only MNO who leases $B$ incurs the investment cost $c_{BS}$.

\begin{figure}[t]
\centering
\includegraphics[angle=0,width=3.4in]{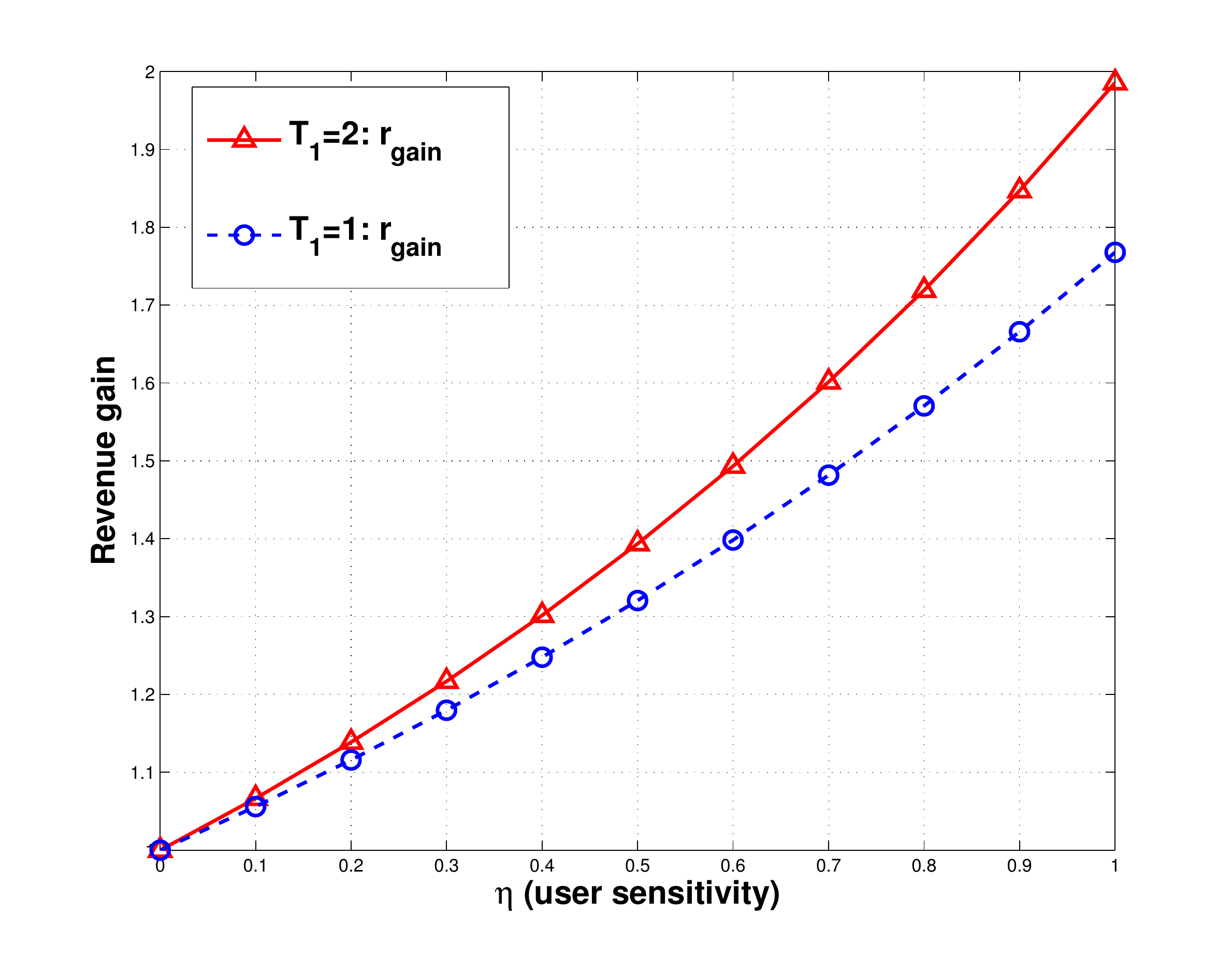}
\caption{Revenue gain as a function of $\eta$ under two different
times ($T_1=1$, $T_1=2$). Other parameters are $u_o=1$,
$\lambda=0.01$ and $T_2=10$.}
\end{figure}
When asymmetric-valued spectrum blocks are allocated to the MNOs,
there is a trade-off between self-interest and spite. To illustrate
this trade-off, we first restrict ourselves to the case where spite
is not present. If MNO $i$ is \textit{self-interested}, his
objective function is as follows. \setlength\arraycolsep{5pt}
\medmuskip=1mu\thinmuskip=1mu\thickmuskip=1mu
\begin{eqnarray}
\Pi _i (b_i ,b_j ) = \left[ {r^A(T_1,T_2)    - b_i } \right] \cdot I_{b_i  \geqslant b_j }  + \pi^{B}(T_1,T_2)   \cdot I_{b_i  < b_j },
\end{eqnarray}
where $I$ is the indicator function and $\pi^{B}(T_1,T_2)=r^{B}(T_1,T_2) - c_B - c_{BS}$ is the profit when leasing $B$.
This case is the standard auction framework in that MNO $i$ maximizes his own profit without considering the other MNO's profit.

In the real world, however, there is some evidence that some MNOs are \textit{completely malicious}.
The German third generation (3G) spectrum license auction in 2000 is a good example \cite{V. Grimm:2002}.
German Telekom kept raising his bid to prevent his competitors from leasing spectrum.
If MNO $i$ is completely malicious, his objective function can be changed as follows.
\setlength\arraycolsep{5pt}
\medmuskip=-1mu\thinmuskip=-1mu\thickmuskip=-1mu
\begin{eqnarray}
\Pi _i (b_i ,b_j ) = \left[ {r^A(T_1,T_2)    - b_i } \right] \cdot I_{b_i  \geqslant b_j } - \left[ {r^A(T_1,T_2)   - b_j } \right] \cdot I_{b_i  < b_j }.
\end{eqnarray}
It means that MNO $i$ gets disutility as much as the profit of MNO $j$ when he loses the auction. The minus term in (22)
implies this factor.

To reflect this strategic concern, our model departs from the standard auction framework
in that each spiteful MNO concerns about maximizing his own profit when he leases $A$, as well as minimizing the weighted difference of his competitor's profit
and his own profit when he leases B.
Combining (21) and (22), we define each MNO's objective function as follows.

\vskip 10pt \noindent {\bf Definition 3.} {\it Assume that two
spiteful MNOs (i.e., $i,j \in \{ 1,2\}$ and $i\neq j$) compete in a
first-price sealed-bid auction. The objective function that each MNO
tries to maximize is given by:
\setlength\arraycolsep{-0.8pt}
\medmuskip=0.5mu\thinmuskip=0.5mu\thickmuskip=0mu
\begin{eqnarray}
\Pi _i (b_i ,b_j )=&{\rm{ }}{\rm{ }}&\left[ {r^A(T_1,T_2)    - b_i } \right] \cdot I_{b_i  \geqslant b_j } \nonumber\\
&+&\left[ {(1 - \alpha _i )\pi ^B (T_1 ,T_2 ) - \alpha _i (r^A (T_1 ,T_2 ) - b_j )} \right] \cdot I_{b_i  < b_j } \nonumber\\
\end{eqnarray}
\medmuskip=3mu\thinmuskip=3mu\thickmuskip=2mu
where $I$ is the indicator function, $\pi^B(T_1,T_2)=r^B(T_1,T_2) - c_B - c_{BS}$ is the MNO's profit when leasing $B$, and $\alpha _i  \in [0,1]$ is a parameter called the spite (or competition) coefficient.
}

As noted, MNO $i$ is self-interested and only tries to maximize his own profit when $\alpha_i =0$. When $\alpha_i =1$,
MNO $i$ is completely malicious and only attempts to obtain more market share by forcing MNO $j$ to lease the less-valued spectrum block.
For given $\alpha _i  \in [0,1]$ and $\alpha _j  \in [0,1]$, we can derive the optimal bidding strategies
that maximize the objective function in \textrm{D}efinition 3 as follows.

\vskip 10pt \noindent {\bf Proposition 3.}  {\it In a first-price
sealed-bid auction, the optimal bidding strategy for a spiteful MNO
$i, j \in\{1,2\}$ and $i\ne j$ is: \setlength\arraycolsep{0pt}
\medmuskip=0.5mu\thinmuskip=0.5mu\thickmuskip=0.5mu
\begin{eqnarray}
b_i^\ast={{(1+\alpha_i)r^A (T_1 ,T_2 ) - (1-\alpha_i)\pi ^B (T_1 ,T_2 ) + c_A } \over {2 + \alpha _i }}, \nonumber
\end{eqnarray}
\begin{eqnarray}
b_j^\ast={{(1+\alpha_j)r^A (T_1 ,T_2 ) - (1-\alpha_j)\pi ^B (T_1 ,T_2 ) + c_A } \over {2 + \alpha _j }}.
\end{eqnarray}
} \vskip 10pt \noindent {\bf Proof.}
Without loss of generality,
suppose that MNO $i$ knows his bid $b_i$.
Further, we assume that MNO $i$ infer that the bidding strategy of MNO $j$ on $A$ is drawn uniformly and independently from $\left[ {c_A ,r^A (T_1,T_2)  } \right]$.
The MNO $i$'s optimization problem is to choose $b_i$ to maximize the expectation of
\setlength\arraycolsep{-1pt}
\medmuskip=0.5mu\thinmuskip=0.5mu\thickmuskip=0.5mu
\begin{eqnarray}
&&E_{b_i } (\Pi _i ) = \int\limits_{c_A }^{b_i } {\left[ {r^A (T_1 ,T_2)  - b_i } \right]} {\rm{ }}f(b_j )db_j  \nonumber \\ &&+  \int\limits_{b_i }^{r^A (T_1 ,T_2 )} {\left[ {(1 - \alpha _i )(\pi ^B (T_1 ,T_2 )) - \alpha _i (r^A (T_1 ,T_2 ) - b_j )} \right]} {\rm{ }}f(b_j )db_j
. \nonumber \\
\end{eqnarray}
Differentiating Equation (25) with respect to $b_i$, setting the result to zero and multiplying by $
r_A^ *   - c_A$ give
\setlength\arraycolsep{1pt}
\medmuskip=5mu\thinmuskip=5mu\thickmuskip=5mu
\begin{eqnarray}
\frac{{\partial E_{b_i } (\Pi _i )}}
{{\partial b_i }} = &&(1+\alpha_i)r^A(T_1,T_2) -(1-\alpha_i)\pi^B(T_1,T_2) \nonumber
\\&&+ c_A  - (2 +\alpha _i )b_i  = 0. \nonumber
\end{eqnarray}
Since the same analysis can be applied to the MNO $j$, the proof is complete.
\hfill $\blacksquare$

\textrm{P}roposition 3 states that the MNOs' equilibrium bidding strategies.
Intuitively, the more spiteful the MNO is, the more aggressively the MNO tends to bid.
For consistency, we assume that $\alpha _i  > \alpha _j$. Then we can now calculate MNO $i$'s profit and MNO $j$'s profit as follows
\medmuskip=2mu\thinmuskip=2mu\thickmuskip=2mu
\begin{eqnarray}
\pi _i^ *   = {{r^A (T_1 ,T_2 ) - (1-\alpha_i)\pi ^B (T_1 ,T_2 )+c_A} \over {2 + \alpha _i }},\;  \pi _j^ *   = \pi ^B (T_1 ,T_2 ), \nonumber \\
\end{eqnarray}
where $\pi_i^*$ is calculated by substraction of the bidding price $b_i^*$ of (24) from $r^A(T_1,T_2)$ of (19).
\begin{figure}[t]
\centering
\includegraphics[angle=0,width=3.4in]{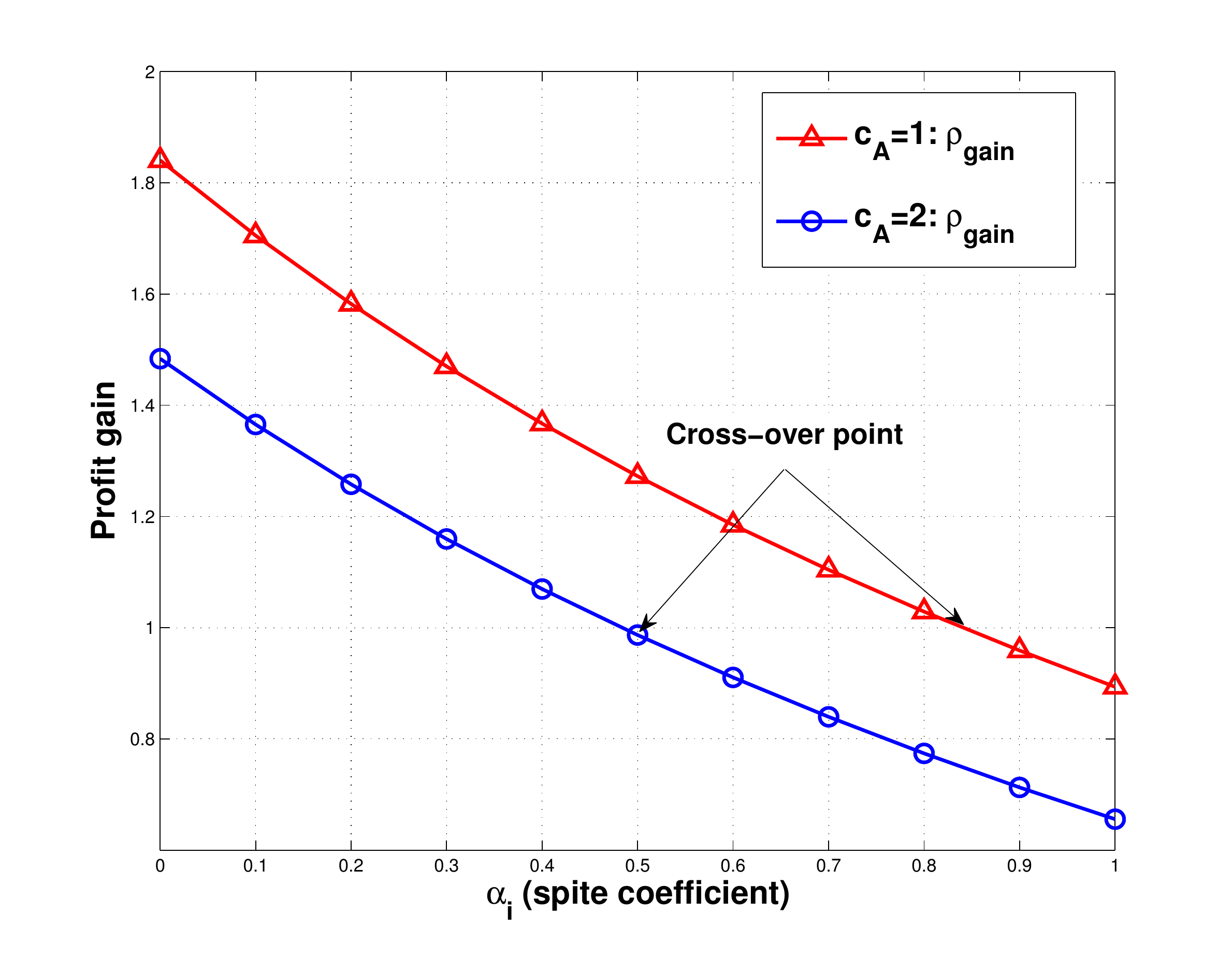}
\caption{Profit gain as a function of $\alpha_i$ under two different
costs ($c_A=1$, $c_A=2$). Other parameters are $u_o=1$,
$\lambda=0.01$, $\eta=0.6$, $T_1=1$, $T_2=10$, $c_B=1$, and
$c_{BS}=1$.}
\end{figure}
\begin{figure}[t]
\centering
\includegraphics[angle=0,width=3.4in]{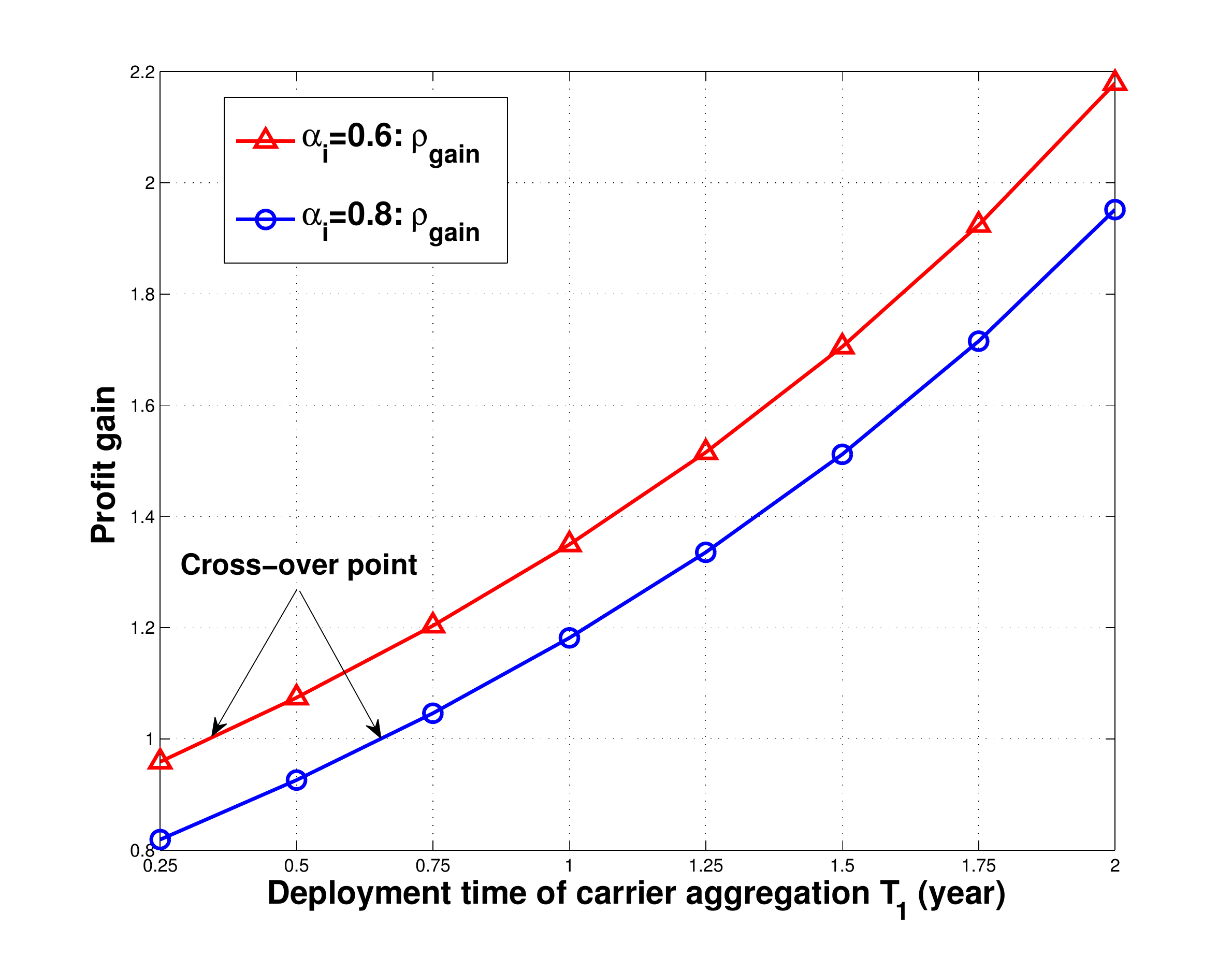}
\caption{Profit gain as a function of $T_1$ under two different
spite coefficients ($\alpha_i=0.6$, $\alpha_i=0.8$). Other
parameters are $u_o=1$,$\lambda=0.01$, $\eta=0.6$, $T_2=10$,
$c_A=2$, $c_B=1$, and $c_{BS}=1$.} \vspace{-10pt}
\end{figure}

To get some insight into the properties of the MNOs' equilibrium profits, let us define
$\rho_{gain}  = \frac{\pi_i^{\ast}}{\pi_j^{\ast}}$\footnote{The gain $\rho_{gain}$ is different from $r_{gain}$ of (19) where $r_{gain}(T_1,T_2)$
is the revenue gain from $A$ relative to $B$ without considering any cost.},
which can be interpreted as the profit gain from $A$ relative $B$.
When $\rho_{gain}>1$, the profit of MNO $i$ is higher than that of MNO $j$. It implies that MNO $i$
could gain a competitive advantage over MNO $j$ in both market share and profit.
When $\rho<1$, the situation is reversed.
MNO $j$ could take the lead in the profit despite losing some market share to MNO $i$.

If the role of the government is to ensure fairness in two MNOs' profits, the government may devise two different schemes:
setting appropriate reserve prices and imposing limits on the timing of the double-speed LTE services.
According to the Ofcom report, setting the reserve prices closer to market value might be appropriate \cite{dotcon_12}.
It indicates that the government set $c_A$ and $c_B$ by estimating the value asymmetries between spectrum blocks $A$ and $B$ (i.e., $r^A(T_1,T_2)-r^B(T_1,T_2)$)
and the spite coefficient $\alpha_i$.
\textrm{F}ig. 6 shows the profit gain as a function of $\alpha_i$ under two different reserve prices for $A$ (i.e., $c_A=1$, $c_A=2$).
For example, if $\alpha_i=0.5$, the government should set the reserve prices $c_A=2$, $c_B=1$.
On the other hand, the government should set the reserve prices $c_A=1$, $c_B=1$ when $\alpha_i=0.85$.

Besides setting appropriate reserve prices, the government can impose limits on the timing of the double-speed LTE service.
In South Korea, for instance, Korea Telecom (KT) who acquired the continuous spectrum spectrum
is allowed to start its double-speed LTE service on metropolitan areas immediately in September 2013, other major cities
staring next March, and nation-wide coverage starting next July \cite{Yoshio_13}. This scheme implies to reduces $T_1$ by limiting
the timing of the double-speed LTE service to the MNO who acquires spectrum block $A$.
\textrm{F}ig. 7 shows the profit gain as a function of $T_1$ under two different spite coefficients (i.e., $\alpha_i=0.6$, $\alpha_i=0.8$).

\section{Conclusion}
In this paper, we study bidding and pricing
competition between two spiteful MNOs with considering their existing spectrum
holdings. We develop an analytical framework to investigate the
interactions between two MNOs and users as a three-stage dynamic game.
Using backward induction, we characterize the dynamic game's
equilibria. From this, we show the asymmetric pricing structure and different market share between two MNO.
Perhaps counter-intuitively, our results show that the MNO who acquires the less-valued spectrum block
always lowers his price despite providing double-speed LTE service to users.
We also show that the MNO who acquires the high-valued spectrum block, despite charging a higher price,
still achieves more market share than the other MNO.
We further show that the competition between two MNOs leads to some loss of their
revenues. With the example of South Korea, we investigate the cross-over point at which two MNOs' profits are switched,
which serves as the benchmark of practical auction designs.

Results of this paper can be extended in several directions.
Extending this work, it would be useful to propose some methodologies for setting reserve prices \cite{S. M. Yu:13}, \cite{S. Y. Jung_13}.
Second, we could consider an oligopoly market where multiple MNOs initially have different market share before spectrum allocation,
where our current research is heading.

\end{document}